\documentclass[12pt]{iopart}

\usepackage{graphicx}
\usepackage{cite}
\usepackage{color}

\begin{document}
\bibliographystyle{iopart-num}

\title[$\gamma$-CAST]{A computationally assisted spectroscopic technique to measure secondary electron emission coefficients in radio frequency plasmas}

\author{M. Daksha$^1$, B. Berger$^{1,2}$, E. Schuengel$^1$, I. Korolov$^3$, A. Derzsi$^3$, M. Koepke$^1$, Z. Donk\'o$^3$, J. Schulze$^1$}
\address{$^1$Department of Physics, West Virginia University, Morgantown, WV 26506, USA. \\ 
$^2$ Department of Electrical Engineering, Ruhr University Bochum, 44780 Bochum, Germany\\
$^3$Institute for Solid State Physics and Optics, Wigner Research Centre for Physics,
Hungarian Academy of Sciences, 1121 Budapest, Konkoly-Thege Mikl\'os str. 29-33, Hungary.}
\ead{manudaksha@gmail.com}

\begin{abstract}
A Computationally Assisted Spectroscopic Technique to measure secondary electron emission coefficients ($\gamma$-CAST) in capacitively-coupled radio-frequency plasmas is proposed. This non-intrusive, sensitive diagnostic is based on a combination of Phase Resolved Optical Emission Spectroscopy and particle-based kinetic simulations. In such plasmas (under most conditions in electropositive gases) the spatio-temporally resolved electron-impact excitation/ionization rate features two distinct maxima adjacent to each electrode at different times within each RF period. While one maximum is the consequence of the energy gain of electrons due to sheath expansion, the second maximum is produced by secondary electrons accelerated towards the plasma bulk by the sheath electric field at the time of maximum voltage drop across the adjacent sheath. Due to these different excitation/ionization mechanisms, the ratio of the intensities of these maxima is very sensitive to the secondary electron emission coefficient $\gamma$. This sensitvity, in turn, allows $\gamma$ to be determined by comparing experimental excitation profiles and simulation data obtained with various $\gamma$-coefficients. The diagnostic, tested here in a geometrically symmetric argon discharge, yields an effective secondary electron emission coefficient of $\gamma = 0.066 \pm 0.01$ for stainless steel electrodes.


\end{abstract}

\submitto{J. Phys. D: Appl. Phys.}

\maketitle

Low temperature radio frequency (RF) plasmas are frequently used for a variety of technological processes where plasma-surface interactions on microscopic scales are utilized in a controlled manner in the manufacturing of different high-tech products \cite{LiebermanBook,MakabeBook,ChabertBook}. While the plasma alters its own boundary surfaces, depending on the shape of the flux-energy distribution functions of different particle species (electrons, ions, neutral radicals), the surface also affects the plasma via particle reflection, absorption, and generation. One of the most important plasma-surface interactions that can strongly affect the electron power absorption dynamics, the plasma density, sheath width, and other plasma parameters is the {\it emission of secondary electrons} (``$\gamma$-process'') induced by the bombardment of the electrode surfaces by different species from the plasma (ions, neutrals, electrons, or photons)\cite{Gamma1,Belenguer1990,Derzsi_Gamma,Braginsky_Gamma,Schulze_Gamma,Booth_Gamma,Boehm_Gamma,Liu_Gamma,Brz_Gamma,
Bogaerts_Gamma3,Gans_Surf}. For instance, as the $\gamma$-coefficient increases due to a change of the electrode material or its conditions in electropositive plasmas operated at pressures above $\sim$50~Pa and at driving voltage amplitudes above $\sim$100~V, an electron heating mode transition can be induced from the $\alpha$-mode, where ionization by electrons accelerated by the expanding sheaths dominates, to the $\gamma$-mode, where ionization due to secondary electrons is most important. This mode transition is typically accompanied by a drastic increase of the plasma density \cite{Schulze_Gamma,Schulze_PRL}.

Depending on the discharge conditions, the importance of secondary electron emission, due to the impact of the different species mentioned above, may vary significantly \cite{Gamma1}. Moreover, the probability of emitting a secondary electron per incident particle depends on the impact energy and angle, as well as on the surface material and its conditions \cite{Gamma1,Boehm_Gamma,GammaE_Angle,GammaI_Material1,GammaI_Material2}. The joint action of all the different species can be expressed by an {\it effective} secondary electron emission coefficient that corresponds to the ratio of the emitted secondary electron and the incident ion fluxes at the surface, implicitly including the contributions of the other species to the secondary electron emission \cite{Gamma1,EffGamma_Donko1,EffGamma_Donko2}.

Primarily due to the lack of detailed data, this complex picture is usually simplified in discharge models. Commonly, surface processes are either completely neglected or only ion-induced secondary electron emission at a constant probability (typically {\it guessed} to be $\gamma \sim 0.1$) is included. Any dependencies on the incident particle energy and angle, as well as on the surface material and its conditions are typically disregarded, in spite of recent Particle-in-Cell (PIC) simulations of capacitive RF plasmas that show drastic effects of including a more realistic implementation of secondary electron emission \cite{Derzsi_Gamma}.

We note that secondary electron emission coefficients reported in the literature are conventionally measured by particle-beam experiments under high vacuum conditions in the absence of plasma and with ultra-clean surfaces \cite{Gamma1,Gamma2,Gamma3}. Unfortunately, these values are not directly applicable for the description of gas discharges, as the presence of a low temperature plasma can strongly affect the surface material, e.g., via deposition or etching (which are the main processes in surface treatment), and can change its secondary electron emission coefficients for the various incident particle species. Therefore, an {\it in-situ} determination of the secondary electron emission coefficient would be highly valuable. 

Here, we propose a novel non-intrusive and in-situ Computationally Assisted Spectroscopic Technique to measure secondary electron emission coefficients ($\gamma$-CAST) based on a combination of experimental Phase Resolved Optical Emission Spectroscopy (PROES) and self-consistent numerical plasma simulations. This diagnostic is applicable to any surface material exposed to a capacitive RF plasma and potentially also to other types of discharges. It is based on measuring,  space and time resolved within the RF period, the electron impact excitation rate from the ground state into a specific excited state of neutral gas atoms in the reactor \cite{Schulze_PROES}. Adjacent to each electrode and for electropositive gases, this measurement typically yields two maxima at distinct times within the RF period -- one caused by the energy gain of the electrons at sheath expansion and another caused by the acceleration and collisional multiplication of secondary electrons during their flight through the space charge sheaths. The simulations are executed for a sequence of $\gamma$ coefficient values (used as an input parameter) under conditions identical to the experiments with the exception of a small Ne admixture used in the experiment, but not included in the simulation (the possible effects of this will be discussed later). As the ratio of the intensities of the two characteristic maxima of the excitation rates depends on the choice of $\gamma$ in the simulation, good agreement between the PROES measurements and the simulations is found only for a specific choice of $\gamma$. In this way the effective secondary electron emission coefficient can be determined accurately.

\begin{figure}[h!]
\begin{center}
\includegraphics[width=0.8\textwidth]{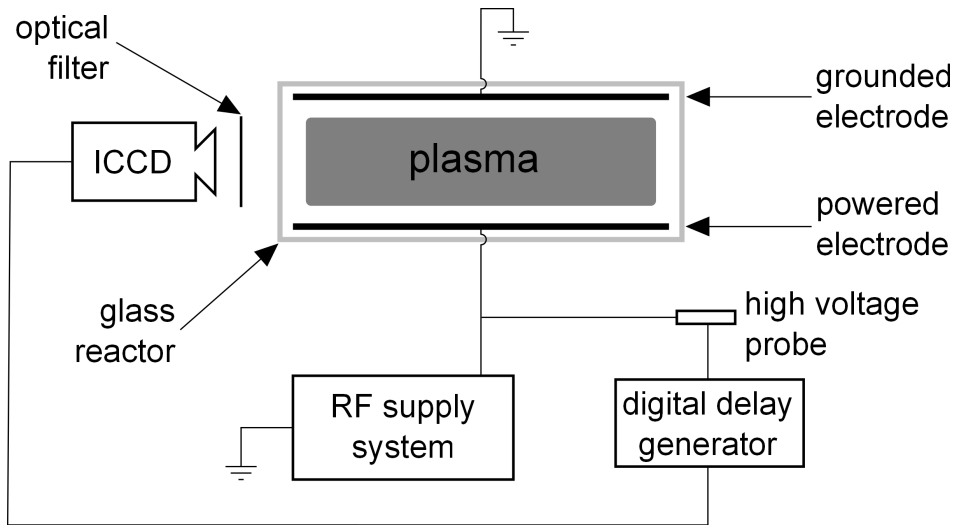}
\caption{Sketch of the experimental setup including all diagnostics.}
\label{fig1}
\end{center}
\end{figure}

Figure \ref{fig1} shows the experimental setup. The reactor consists of two plane parallel and circular electrodes made of 316 stainless steel and separated by a gap of 2.5 cm inside a glass reactor. The diameter of both electrodes is 10 cm. One electrode is driven by a single frequency driving voltage waveform, $\tilde{\phi}(t) = -\phi_0 \cos{(2 \pi f t)}$, via an impedance matching network, while the other electrode is grounded. Here, $f = 13.56$ MHz is the driving frequency and $\phi_0$ is the driving voltage amplitude. The discharge is operated in argon gas with neon added as a tracer gas for PROES \cite{Schulze_PROES2}. 

During the course of the measurements, different admixture ratios were tested. A low Ne concentration results in an unacceptable signal to noise ratio due to the weak intensity of the observed Ne emission line, while a high Ne concentration was found to affect the electron impact excitation dynamics. A 10\% Ne addition was found to be an acceptable compromise in this respect, since the spatio-temporal PROES data were practically identical to those obtained with only 4\% of Ne, but the signal-to-noise ratio was much better. 

We apply a constant driving power of 20 W over a range of neutral gas pressures, between 75 Pa and 175 Pa, for which the driving voltage and the RF current adjust to maintain the constant power constraint. A fast high-voltage probe is used to measure, time resolved within the RF period, the voltage drop across the discharge directly adjacent to the powered electrode. In this way, $\phi_0$ and the DC self-bias are measured. As the entire reactor wall is made of glass and only a single driving frequency is used, the plasma is completely symmetric and the measured DC self-bias is less than 1 \% of the driving voltage amplitude under all conditions investigated. The PROES measurements are carried out employing a fast ICCD camera equipped with an interference filter. The camera is synchronized with the RF driving voltage waveform via a delay generator. The plasma emission at 585.2 nm originating from the Ne2p$_1$ state is measured with 2 ns time resolution and about 0.15 mm spatial resolution in the direction perpendicular to the electrodes. This neon state is used due to its short lifetime of about 15 ns, its high excitation threshold of about 19 eV, and its simple population dynamics \cite{Schulze_PROES}. A collisional-radiative model is used to calculate the electron impact excitation rate from the ground state into this excited state with the same temporal and spatial resolution from the measured emission. Due to the high excitation threshold, the obtained excitation rate probes the electron impact ionization rate of argon \cite{Schulze_PROES}. 

The discharge symmetry is crucial, since, in the frame of the proposed diagnostic to measure $\gamma$-coefficients, the experimentally determined spatio-temporal excitation rate is compared to simulation results obtained from a benchmarked \cite{PIC_Benchmark} 1d3v electrostatic PIC simulation code complemented with Monte-Carlo treatment of collision processes (PIC/MCC) \cite{Donko_PIC}, wherein a symmetric discharge configuration is inherently assumed. This simulation is used to obtain, space and time resolved, the electron impact ionization rate of argon. The  cross section sets for electron-neutral and ion-neutral collision processes are taken from the literature \cite{Gamma1,PIC_CS}. The measured driving voltage amplitude is used as an input parameter for the simulation at each pressure investigated. In the simulation, the neutral gas temperature is set to 350 K. The computational grid comprises 1000 points and the RF period is divided into 15~000 time steps in order to fulfill all stability criteria \cite{Donko_PIC}. The electron reflection probability at the electrodes is set to 20 \% \cite{Kollath,Bronshtein}. The (effective) $\gamma$-coefficient is an input parameter and the simulations have been run for a sequence of $\gamma$-coefficients at each pressure. The results of these different simulation runs are compared to the experiment. Good agreement is only found for a specific choice of $\gamma$. In this way $\gamma$ is determined uniquely. We note that the electron reflection probability of the electrode material is also less known, when the electrode is exposed to a plasma \cite{Korolov_Refl,Bronold_Refl,Koepke_Refl}. Therefore, we quantitatively profiled the influence of this parameter on the simulation results. This profile assesment concluded that a higher electron reflection probability leads to a higher plasma density, as well as a higher ion flux to the electrodes, but it affects the intensities of both maxima in the electron impact excitation rate in a similar way, i.e., its influence on the determination of the $\gamma$-coefficient via $\gamma$-CAST is negligible. In the simulations the presence of the Ne admixture was not taken into account to simplify the computational aspect of this diagnostic. This is justified, because varying the relative Ne admixture between 4 \% and 10 \% in the experiment was found to have a negligible effect on the measured ratio of the intensities of both maxima, i.e. the uncertainty in determining $\gamma$ due to this Ne admixture is much smaller than the estimated total uncertainty of $\pm 0.01$ (see discussion below).


\begin{figure}[h!]
\begin{center}
\includegraphics[width=1\textwidth]{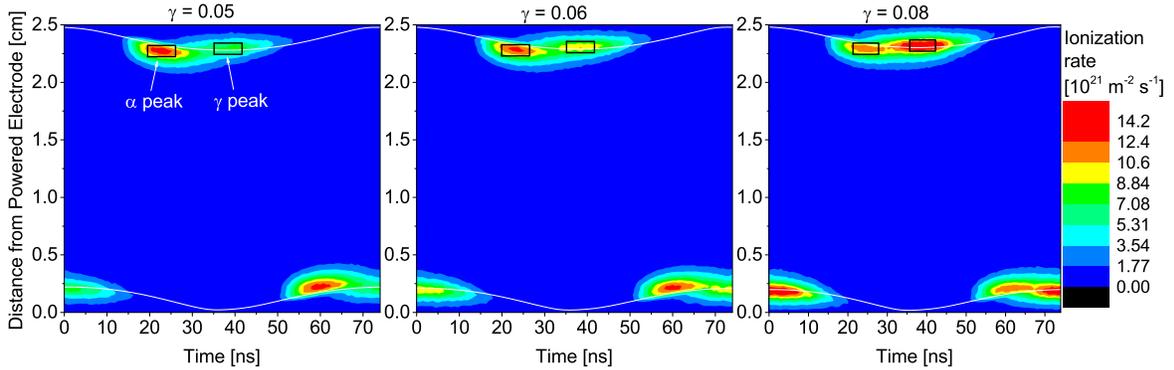}
\caption{Spatio-temporal plots of the ionization rate obtained from the simulations at 150 Pa and $\phi_0$ = 244 V, for different effective $\gamma$-coefficients. The rectangles indicate the regions of interest (ROI) around the ionization maxima (see text). Averages over these ROIs are used for the further data analysis. The white lines indicate the width of the sheath adjacent to each electrode \cite{Brinkmann}. The units are $10^{21}$ m$^{-2}$ s$^{-1}$.}
\label{fig2}
\end{center}
\end{figure}

Figure \ref{fig2} shows representative examples of spatio-temporal plots of the ionization rate obtained from the simulation at 150 Pa, for three values of the secondary electron emission coefficient: $\gamma$ = 0.05, 0.06, and 0.08. These plots are obtained by acquiring data over 7500 consecutive RF periods (after convergence of the simulation). Adjacent to each electrode, we observe two maxima. The first maximum is caused by the energy gain of electrons upon sheath expansion ($\alpha$-mechanism of the energy gain, at $t \approx 22$ ns, at the top electrode), while the second maximum is caused by secondary electrons emitted from the electrode, which are accelerated towards the plasma bulk by the high sheath electric field and are multiplied inside the sheaths by collisions ($t \approx 37$ ns, at the top electrode). The second maximum is observed around the time of maximum local sheath voltage (maximum sheath extension) within the RF period. The same maxima are observed at the bottom electrode half an RF period later. At the top electrode the maxima are marked by rectangles, which serve as regions of interest (ROI). The width and height of both ROIs are chosen by finding the positions and times where/when the intensity decays to 80 \% of the peaks. By systematically changing this limiting value from 80 \% to 90 \% the specific choice of the limiting value within this interval was found to have a negligible effect on the results. The typical dimensions of the ROIs defined this way are $\sim$ 10 ns and $\sim$ 1.5 mm, respectively. For each peak, the intensity is averaged over the respective ROI, resulting in averaged intensities for the $\alpha$- and the $\gamma$-maximum, $I_{\alpha}$ and $I_{\gamma}$. Figure \ref{fig2} reveals the strong sensitivity of the spatio-temporal ionization dynamics on the choice of $\gamma$ in the simulation under these conditions. Clearly, the ratio $I_{\gamma}/I_{\alpha}$ increases as a function of $\gamma$ because more secondary electrons are generated at the electrode.  Alternatively fixed dimensions of the ROIs could be defined for all conditions investigated. However, this will be critical, if these fixed dimensions are chosen in a way that the ROI around one maximum includes regions of lower relative intensity compared to the ROI around the other maximum. Then, the choice of the fixed dimensions would affect the value of $\gamma$ determined by this diagnostic. Similarly a ROI could be chosen to be too large so that parts of other maxima are included. Both effects are critical, since the size of the $\alpha$- and $\gamma$-maxima change as a function of external control parameters such as the pressure. Using the relative 80 \% - criterion avoids these problems. As we have verified with our own data, as long as spatially-fixed dimensions of the ROIs are chosen so that these problems are avoided, the quantitative and qualitative results of $\gamma$-CAST are insensitive to the specific choice of the ROI boundary.

\begin{figure}[h!]
\begin{center}
\includegraphics[width=1\textwidth]{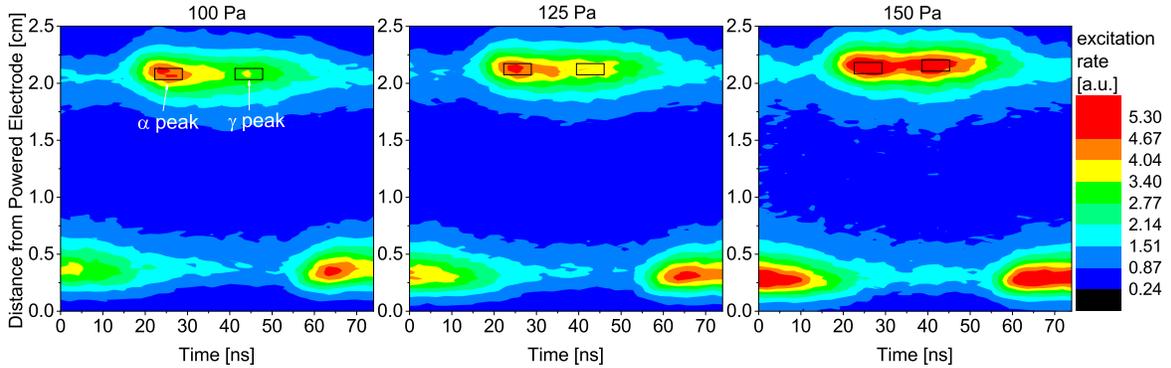}
\caption{Spatio-temporal plots of the electron impact excitation rate from the ground state into the Ne2p$_1$-state obtained experimentally by PROES measurements at different pressures. As the power is kept constant, the driving voltages are different at each pressure (266 V at 100 Pa, 256 V at 125 Pa, and 244 V at 150 Pa).}
\label{fig3}
\end{center}
\end{figure}

\begin{figure}[h!]
\begin{center}
\includegraphics[width=0.7\textwidth]{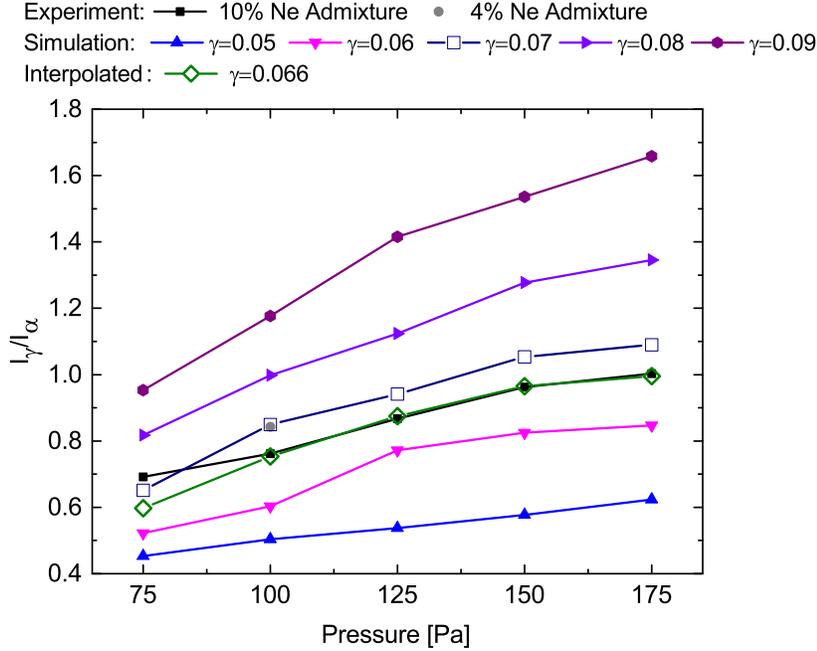}
\caption{Ratio of the averaged intensities of the maxima caused by secondary electrons, $I_{\gamma}$, and by sheath expansion heating, $I_{\alpha}$, obtained from the experiment with 10 $\%$ Ne admixture and the simulation as a function of pressure. The effect of the relative Ne admixture in the experiment on $I_{\gamma}/I_{\alpha}$ is illustrated at 100 Pa (grey solid dot, 4 \% Ne admixture).}
\label{fig4}
\end{center}
\end{figure}


Similar plots of the electron impact excitation rate from the ground state into the Ne2p$_1$-state are obtained experimentally by PROES as a function of pressure. As the driving power is kept constant, the driving voltages are different at each pressure (278 V at 75 Pa, 266 V at 100 Pa, 256 V at 125 Pa, 244 V at 150 Pa, and 234 V at 175 Pa). Representative examples of such plots obtained at 100 Pa, 125 Pa, and 150 Pa are shown in figure \ref{fig3}. Due to the high energy threshold, this excitation rate probes the ionization rate of argon \cite{Schulze_PROES2}. Similar to the simulation results shown in figure \ref{fig2}, two distinct maxima at each electrode can be identified at different times within the RF period. ROIs (rectangles), with dimensions determined in the same way as done for the simulation data, are centered around the maximum intensities. In this way, the intensity ratio $I_{\gamma}/I_{\alpha}$ is obtained experimentally for each pressure. The increasing pressure induces an $\alpha$- to $\gamma$-mode transition as this ratio increases. 

In order to determine the unknown effective $\gamma$-coefficient, the intensity ratios $I_{\gamma}/I_{\alpha}$ obtained from the experimental PROES measurements and the PIC simulations are plotted as a function of the neutral gas pressure in figure \ref{fig4}. In the simulation, $\gamma$ is varied, so that figure \ref{fig4} displays separate lines obtained from the simulation for different secondary electron emission coefficients. The best agreement between the experimental intensity ratio and the simulation data appears for $\gamma \approx 0.066$ (for the stainless steel electrodes used here). This value is obtained based on a linear interpolation between the simulation data for $\gamma =$ 0.06 and 0.07 at each pressure. Matching experiment and simulation in this way the effective $\gamma$-coefficient is determined from PROES measurements and PIC/MCC simulations in-situ and non-intrusively. In figure \ref{fig4} the experimental data obtained with 10 \% Ne admixture yields a line, which is parallel to the simulation data obtained for $\gamma = 0.07$ for pressures between 100 Pa and 175 Pa. At 75 Pa the intensity of the maximum caused by the secondary electrons becomes very weak in the experiment and the simulation, but the result for $\gamma$ is still within the uncertainty of $\gamma$-CAST ($\pm 0.01$). In principle applying this diagnostic to a single set of conditions of interest is sufficient to obtain a realistic effective secondary electron emission coefficient for these conditions quickly.

\begin{figure}[h!]
\begin{center}
\includegraphics[width=0.6\textwidth]{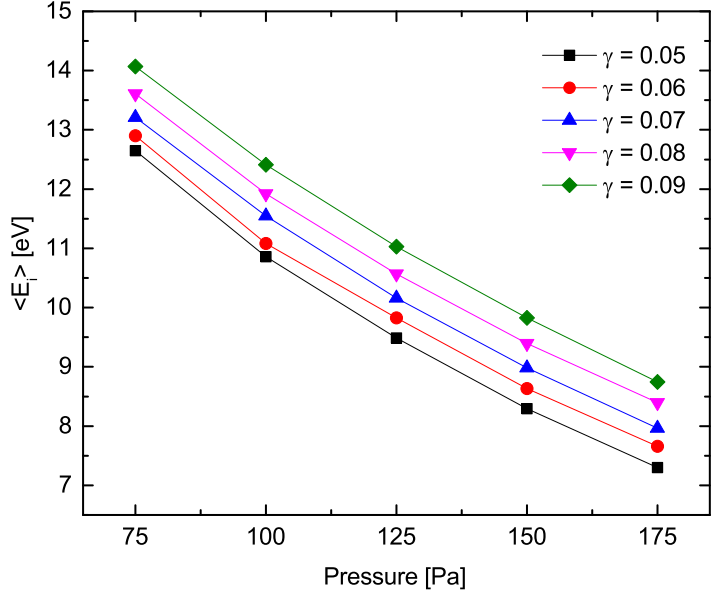}
\caption{Mean ion energy at the electrodes as a function of pressure obtained from the simulation for different effective $\gamma$-coefficients. Measured driving voltage amplitudes are used as input parameters for the simulation at each pressure.}
\label{fig5}
\end{center}
\end{figure}

Figure \ref{fig5} shows the mean ion energy at the electrodes as a function of pressure for all conditions investigated based on the simulation. Under all conditions the mean ion energy is relatively low, i.e. between about 7 eV and 14 eV. The mean ion energy decreases as a function of pressure, because the sheaths get more collisional. For a given pressure it increases as a function of $\gamma$, since the sheath width decreases due to the higher plasma density and, therefore, gets less collisional. According to the analysis of the contributions of different species to secondary electron emission made in \cite{Gamma1}, ions are expected to play the major role at our conditions. As the ion energies are comparatively low, we do not expect a significant contribution by fast atoms that are created by ion-neutral charge-exchange collisions inside the sheaths. For Ar$^+$ ions, a secondary yield of $\gamma \approx 0.07$ is given in \cite{Gamma1} for low-energy ion impact at ``clean'' metal electrode surfaces. This value is in very good agreement with that found in the present work. The electrode surfaces in our experiment are likely to qualify to be ``clean'' (see \cite{Gamma1} for details of the terminology) because they have been exposed to energetic ion bombardment by operating a low pressure discharge at high driving voltages prior to the reported measurements, and during the measurements the surfaces are only exposed to inert noble gases. In reference \cite{Gamma1} analytical equations are provided that allow to calculate the secondary electron emission coefficient due to ion and fast atom impact on clean metal surfaces as a function of the incident particle energy (equations B10 and B12 in \cite{Gamma1}). According to these equations fast atoms below 500 eV do not cause secondary electron emission and the ion induced $\gamma$-coefficient is 0.07 for ion energies below 500 eV. Based on the ion energy distributions at the electrodes obtained from the PIC/MCC simulation under the conditions investigated here, the incident ion energies are limited to values much lower than 500 eV. Therefore, also the energies of fast atoms are limited to values much lower than 500 eV, since these particles are maily created by charge-exchange collisions between ions and neutrals. Therefore, based on previous measurements of $\gamma$-coefficients we expect a value of $\gamma = 0.07$, which is in very good agreement with the value for $\gamma$ obtained by $\gamma$-CAST.

As mentioned above, several factors may influence the determination of the effective secondary electron emission coefficient. These are: (i) the statistical error of the PROES measurements and the simulation data, (ii) the use of Ne admixture in the experiment, (iii) the unknown reflection coefficient of electrons at the electrodes used as an input parameter in the simulation, and (iv) the systematic, although somewhat arbitrary, determination of the ROIs. The statistical error due to statistical changes of the intensities of both maxima in consecutive measurements/simulations under identical conditions was found to result in a maximum uncertainty of $I_\gamma/I_\alpha$ of 0.05. The effect of Ne is twofold: it may influence the electron kinetics due to different cross sections for collisions; Ne$^+$ ions may also contribute to secondary electron emission. Therefore, the effect of changing the relative Ne admixture was tested systematically in the experiment and an admixture of 10 \% was found to have only a weak effect on the measured intensity ratio, i.e. the intensity ratios obtained with Ne admixtures of 10 \% or less are the same within the statistical uncertainty of this diagnostic (see figure \ref{fig4}). This is caused by the fact that the admixture is small and, therefore, the electron kinetics are dominated by collisions with Ar atoms and, due to the fact that Ar has significantly lower excitation and ionization energies compared to Ne, a very small number of Ne$^+$ ions is expected to be present in the plasma. The effect of the reflection coefficient has also been tested and a weak influence on the results was found, just like in the tests used in the determination of the ROIs. Based on all of these tests we estimate the uncertainty of the determination of $\gamma$, based on our diagnostic, to be less than $\pm 0.01$.


In conclusion, we developed a novel Computationally Assisted Spectroscopic Technique to measure effective secondary electron emission coefficients non-intrusively and in-situ in capacitive RF plasmas ($\gamma$-CAST). It is based on PROES measurements of the spatio-temporal electron impact excitation rate from the ground state into a particularly chosen excited neon state that probes the ionization rate of the background gas. These measurements show two distinct maxima adjacent to each electrode at different times within the RF period. One maximum is caused by the electron energy gain upon sheath expansion, while the other is caused by secondary electrons emitted from the electrode surface and accelerated towards the plasma bulk by the sheath electric field. The ratio of the intensities of these two maxima was calculated and compared to the results of PIC/MCC simulations of an argon discharge, where $\gamma$ was varied systematically under otherwise identical conditions as used experimentally. The intensity ratios obtained experimentally and from the simulation were compared for a variety of neutral gas pressures and good agreement was found only for a distinct choice of $\gamma$ in the simulation. In this way, the effective secondary electron emission coefficient was determined. Here, this diagnostic was tested for stainless steel electrodes in an argon plasma and $\gamma = 0.066 \pm 0.01$ was found in excellent agreement with previous results for clean metal surfaces \cite{Gamma1}. This diagnostic can, in principle, be applied to any electrode material and gas mixture as long as PROES measurements and PIC/MCC simulations can be performed and as long as the calculated intensity ratio uniquely depends on $\gamma$. Such a systematic investigation of different surface materials exposed to a variety of plasma conditions is, however, beyond the scope of this work, which introduces $\gamma$-CAST as a diagnostic conceptionally. Practically, these constraints restrict this diagnostic to higher pressures, where both peaks can be observed clearly since, at low pressures, no maximum due to secondary electron emission can be observed even for high values of $\gamma$. In contrast to many other methods, this technique takes into account any modification of the surface by the plasma. In principle $\gamma$-CAST can be used for (i) real-time measurements of the effective $\gamma$-coefficient, based on previously obtained simulation results, and (ii) for plasma monitoring to detect process drifts in laboratory as well as industrial environments. It should also be applicable to other plasma sources as long as the maxima of the spatio-temporal electron impact excitation dynamics can be separated in space and time and one of these maxima is sensitive to $\gamma$.

\ack{This work has been supported by a West Virginia Senate Research Grant and the Hungarian Scientific Research Fund through the grant OTKA K-105476. Mark Koepke was supported by NSF grant PHY-1301896.}


\end{document}